\documentclass[showpacs,amsmath,amssymb,aps,showkeys,floatfix,preprint,a4paper]{revtex4}

\usepackage[dvips]{graphicx}
\usepackage{dcolumn}
\usepackage{bm}
\usepackage{epsfig}
\usepackage{amsfonts}
\usepackage{amssymb,amscd}

\def\lsim{\raise0.3ex\hbox{$<$\kern-0.75em\raise-1.1ex\hbox{$\sim$}}}
\def\gsim{\raise0.3ex\hbox{$>$\kern-0.75em\raise-1.1ex\hbox{$\sim$}}}

\def\beq{\begin{equation}}
\def\eeq{\end{equation}}
\def\bea{\begin{eqnarray}}
\def\eea{\end{eqnarray}}
\def\bq{\begin{quote}}
\def\eq{\end{quote}}

\def\gappeq{\mathrel{\rlap {\raise.5ex\hbox{$>$}}
{\lower.5ex\hbox{$\sim$}}}}

\def\lappeq{\mathrel{\rlap{\raise.5ex\hbox{$<$}}
{\lower.5ex\hbox{$\sim$}}}}

\def\Toprel#1\over#2{\mathrel{\mathop{#2}\limits^{#1}}}

\begin{document}


\title{Diffractive vector meson production at large-$t$ in  coherent hadronic interactions at CERN LHC}

\author{V.~P. Gon\c{c}alves}
\email{barros@ufpel.edu.br}
\author{W.~K. Sauter}
\email{werner.sauter@ufpel.edu.br}
\affiliation{High and Medium Energy Group, \\
Instituto de F\'{\i}sica e Matem\'atica, Universidade Federal de Pelotas\\
Caixa Postal 354, CEP 96010-900, Pelotas, RS, Brazil}
\date{\today}

\begin{abstract}
The diffractive vector  meson photoproduction with hadron dissociation in coherent hadron-hadron interactions at LHC energies is studied  assuming  that the color singlet $t$-channel exchange carries large momentum transfer.  We  estimate the rapidity distribution and total cross section for the processes $pp  \rightarrow p V X$ and $PbPb  \rightarrow Pb V X$ ($V = \rho$ or $\Upsilon$) considering the non-forward solution of the BFKL equation  at high energy and large momentum transfer. Our results indicate that the experimental identification of these processes can be feasible at the LHC.

\end{abstract}


\maketitle



In last years several authors have demonstrated that the study of photon - hadron and photon - photon 
interactions  in hadronic collisions is an  alternative way to study the Quantum Chromodynamics (QCD) at high energies \cite{klein_prc,vic_bert,strikman_vec,vicmag_upcs,vicmag_ups,vicmag_rho,motyka_watt,schafer1,cox}. 
The typical examples of processes that can be studied  in these coherent hadron-hadron collisions are the exclusive vector meson production and the inclusive heavy quark production, described by the {reactions} $\gamma h \rightarrow V h$ ($V = \rho, J/\Psi, \Upsilon$) and $\gamma h \rightarrow X Y$ ($X = c\overline{c}, b\overline{b}$),
respectively.  As shown e.g.  in Refs. \cite{vicmag_upcs,vicmag_ups,vicmag_rho}, the study of these processes could valuable information on QCD dynamics. Recently, we have  extended the previous studies for the diffractive $J/\Psi$ photoproduction  with hadron dissociation in the case of
large momentum transfer \cite{vic_sauter} (For related studies see Ref. \cite{fran}). Our goal in this paper is the complement of the Ref. \cite{vic_sauter} for the cases of the $\rho$ and $\Upsilon$ photoproduction. In particular,  the large $t$ $\rho$ photoproduction has been studied in $\gamma p$ interactions in HERA and the experimental data  available \cite{hera_data} allow us to constrain the free parameters and check our theoretical formalism. 


In diffractive vector meson photoproduction with hadron dissociation the { $t$-channel
color singlet} carries large momentum transfer, which  means that the square of the four momentum transferred {across} the associated rapidity gap, -$t$, is large. {Differently} from the diffractive processes studied in Refs. \cite{klein_prc,vic_bert,vicmag_upcs,strikman_vec,motyka_watt,schafer1,cox}, which are characterized by two rapidity gaps with the two hadrons remaining intact, now we still have  two  large rapidity gaps in the detector but  one of the hadrons dissociates. One expects a rapidity gap between the hadron, which emits the photon and remains intact, and the vector meson.
{Concerning the other gap, we expect a vector meson on one side and a
jet on the other, which balances the transverse momentum}. 
The cross section for the diffractive vector meson photoproduction at large momentum transfer  in a coherent  hadron-hadron collision is  given by
\begin{eqnarray}
\frac{d\sigma \,\left[h h \rightarrow   h \otimes V \otimes X \right]}{dy dt} = \omega \frac{dN_{\gamma} (\omega )}{d\omega }\,\frac{d \sigma_{\gamma h \rightarrow V  X}}{dt}\left(\omega \right)\,\,\,,
\label{dsigdydt}
\end{eqnarray}
where $h = p$ or $Pb$, $V = \rho$ or $\Upsilon$,  $\otimes$ {means} the presence of a rapidity gap and the rapidity $y$ of the vector meson produced is directly related to the photon energy $\omega$, i.e. $y\propto \ln \, (2 \omega/m_{V})$. Moreover, $\frac{dN_{\gamma} (\omega )}{d\omega }$ is  the photon spectrum and $\frac{d\sigma}{dt}$ is the differential cross section for the process $\gamma h \rightarrow V X $. The coherence condition {restricts} the photon virtuality to very low values, which implies that for most purposes {the photons} can be considered as real. In what follows we assume that the equivalent photon flux of a nucleus is given  by \cite{upcs}
\begin{eqnarray}
\frac{dN_{\gamma}\,(\omega)}{d\omega}= \frac{2\,Z^2\alpha_{em}}{\pi\,\omega}\, \left[\bar{\eta}\,K_0\,(\bar{\eta})\, K_1\,(\bar{\eta})- \frac{\bar{\eta}^2}{2}\,{\cal{U}}(\bar{\eta}) \right]\,\,\,,
\label{fluxint}
\end{eqnarray}
where
 $\omega$ is the photon energy,  $\gamma_L$ is the Lorentz boost  of a single beam; $K_0(\bar{\eta})$ and  $K_1(\bar{\eta})$ are the
modified Bessel functions.
Moreover, $\bar{\eta}=\omega\,(R_{h_1} + R_{h_2})/\gamma_L$ and  ${\cal{U}}(\bar{\eta}) = K_1^2\,(\bar{\eta})-  K_0^2\,(\bar{\eta})$.
{Eq. (\ref{fluxint})} will be used in our calculations of vector mesons photoproduction in  $PbPb$ collisions. For   proton-proton interactions we assume that the  photon spectrum is given by  
\begin{eqnarray}
\frac{dN_{\gamma}(\omega)}{d\omega} =  \frac{\alpha_{\mathrm{em}}}{2 \pi\, \omega} \left[ 1 + \left(1 -
\frac{2\,\omega}{\sqrt{S_\mathrm{NN}}}\right)^2 \right] 
\left( \ln{\Omega} - \frac{11}{6} + \frac{3}{\Omega}  - \frac{3}{2 \,\Omega^2} + \frac{1}{3 \,\Omega^3} \right) \,\,\,,
\label{eq:photon_spectrum}
\end{eqnarray}
with the notation $\Omega = 1 + [\,(0.71 \,\mathrm{GeV}^2)/Q_{\mathrm{min}}^2\,]$ and $Q_{\mathrm{min}}^2= \omega^2/[\,\gamma_L^2 \,(1-2\,\omega /\sqrt{S_\mathrm{NN}})\,] \approx (\omega/
\gamma_L)^2$.

The differential cross section $d \sigma /dt$  can be calculated considering that for  large momentum transfer the pomeron couples predominantly to individual partons in the hadron \cite{FR,BFLW}. As consequence, the cross section for the photon - hadron interaction can be expressed by the product of the parton level cross section and the parton distribution of the hadron,
\begin{eqnarray}
\frac{d\sigma (\gamma h \rightarrow V X)}{dt dx_j} = f(x_j,|t|)  \, \frac{d\sigma}{dt}(\gamma q \rightarrow V q)\,\,\,,
\label{dsigdtdx}
\end{eqnarray}
where 
\begin{eqnarray}
f(x_j,|t|) = \frac{81}{16} G(x_j,|t|) + \sum_j [ q_j(x_j,|t|) + \bar{q}_j(x_j,|t|)]\,\,\,,
\end{eqnarray} 
and $G(x_j,|t|)$ and $q_j(x_j,|t|)$ are the gluon and quark distribution functions, respectively.  
The struck parton initiates a jet and carries a fraction $x_j$ of the longitudinal momentum of the incoming hadron, which is given by $x_j = -t/(-t + M_X^2 - m^2)$, where $M_X$ is the mass of the {products of the target dissociation} and $m$ is the mass of the target. The minimum value of $x_j$ is calculated considering the experimental cuts on $M_X$.  
Following Refs. \cite{FP,jhep} we calculate $d \sigma /dt$ for the {process} $\gamma h \rightarrow V X$ {by} integrating Eq. (\ref{dsigdtdx}) over $x_j$ in the region $0.01 < x_j < 1$. 
The differential cross-section for the $\gamma q \rightarrow V q$ process
can be expressed as follows
\begin{equation}
\frac{d\sigma (\gamma q \rightarrow V q)}{dt} \; = \;
\frac{16\pi}{81 t^4}
|{\mathcal{F}}(z,\tau)|^{2}\,\,\,,
\label{dsdtgq}
\end{equation}
where $
z = \frac{3\alpha_{s}}{2\pi} \ln (\frac{ s}{\Lambda^{2}})$, $\tau = |t|/(M_{V}^{2}+ Q_{\gamma}^{2})$,
 $M_{V}$ is the mass of the vector meson, $Q_\gamma$ is the photon virtuality and $\Lambda^{2}$ is a characteristic  scale related to $M_V^2$  and $|t|$. In this paper we only consider $Q_\gamma=0$.  
 The BFKL amplitude ${\cal{F}}(z,\tau)$, in the leading logarithmic approximation (LLA) and lowest conformal spin ($n=0$), is given by~\cite{Lipatov}
\begin{equation}
\label{BFKLa}
{\mathcal{F}}(z,\tau)=\frac{t^{2}}{(2 \pi)^{3}}\int d\nu \frac{\nu ^{2}}{(\nu ^{2}+1/4)^{2}}e^{\chi (\nu )z}I_{\nu }^{\gamma V}(Q_{\perp })I^{q q}_{\nu }(Q_{\perp })^{\ast }\,\,\,,
\end{equation}
 where $Q_{\perp}$ is the momentum transferred, $t=-Q_{\perp}^2$ and $\chi (\nu )=4{\mathcal{R}}\mathrm{e} (\psi (1)-\psi (\frac{1}{2}+i\nu ))$ 
is proportional to the BFKL kernel eigenvalues~\cite{Jeff-book} with $\psi(x)$ being the digamma function. The quantities $I_{\nu }^{\gamma V}$ and $I_{\nu }^{qq}$ are given in terms of the BFKL eigenfunctions and impact factors, which describe in the high energy limit the couplings of the external particle pair to the color singlet gluonic ladder (See Ref. \cite{vic_sauter} for details). It is important to emphasize that in LLA, \( \Lambda  \) is arbitrary (but must depend on the scale in the problem, see discussion below) and \( \alpha _{s} \) is a constant.

\begin{figure}[t]
\centerline{\psfig{file=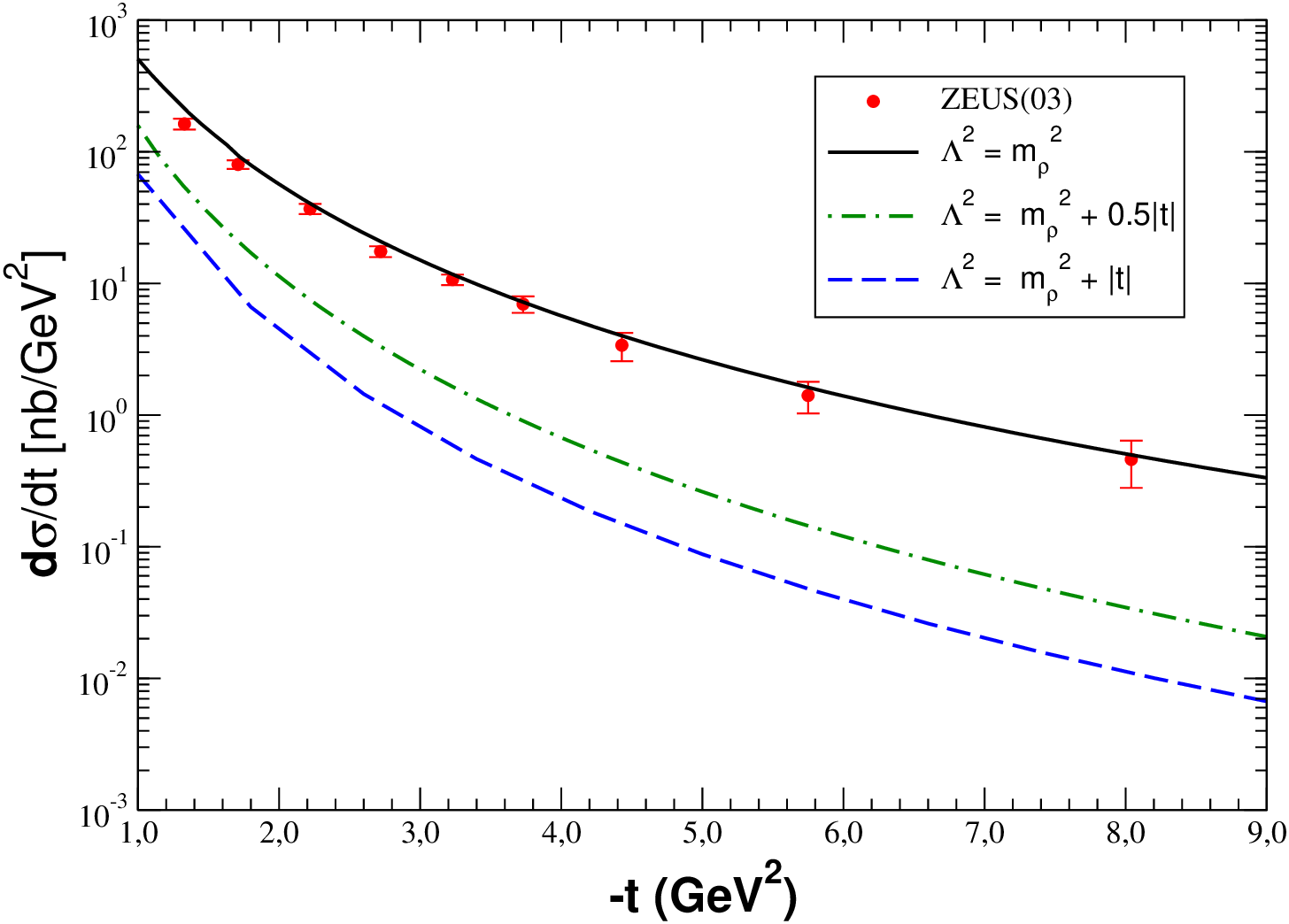,width=100mm}}
 \caption{(Color online) Differential cross section for high-$t$ diffractive $\rho$ photoproduction. Data from Ref. \cite{hera_data}.}
\label{fig1}
\end{figure}




{Let us} start the analysis of our results discussing the diffractive $\rho$ photoproduction in $ep$ collisions at HERA. In particular, it is important clarify our choice for the parameters $\alpha_s$ and $\Lambda$.
{The strong coupling appears in two pieces of the calculations: in the
impact factors, coming from their couplings to the two gluons, and in the definition of the variable $z$, being generated by the gluon coupling inside the gluon ladder} (For details see Ref.\cite{jhep}). As in Ref. \cite{vic_sauter}, we will treat these strong couplings as being identical and we will assume a fixed $\alpha_s$, which is appropriate to the leading logarithmic accuracy. Furthermore, as the cross section is proportional to $\alpha_s^4$, our results are strongly dependent on the choice {of} $\alpha_s$. { Following
Refs. \cite{FP,jhep}, we assume $\alpha_s = 0.21$, with this valued determined
from a fit to the HERA data} (For a detailed discussion see Ref. \cite{FP}).
Similarly, {in the LLA},  $\Lambda$  is arbitrary but must depend on the scale in the problem. In our case we have that in general $\Lambda$ will be a function of $M_V$ and/or $t$. Following Ref. \cite{FP} we {assume} that $\Lambda$ can be expressed by $\Lambda^2 = \beta M_{V}^2 + \gamma |t| $, with $\beta$ and $\gamma$ being free parameters to be fitted by data. 
In Fig. \ref{fig1} we compare our predictions for the differential cross section for $\rho$ production with the HERA data \cite{hera_data}. The curves were obtained using the CTEQ6L parton distributions \cite{cteq6} and considering distinct values for $\beta$ and $\gamma$. A very good agreement with the data is obtained for $\beta = 1.0$ and $\gamma = 0$, which are the values that we will use in our calculations for $\rho$ production.

\begin{figure}[t]
\begin{tabular}{cc}
\includegraphics[scale=0.15] {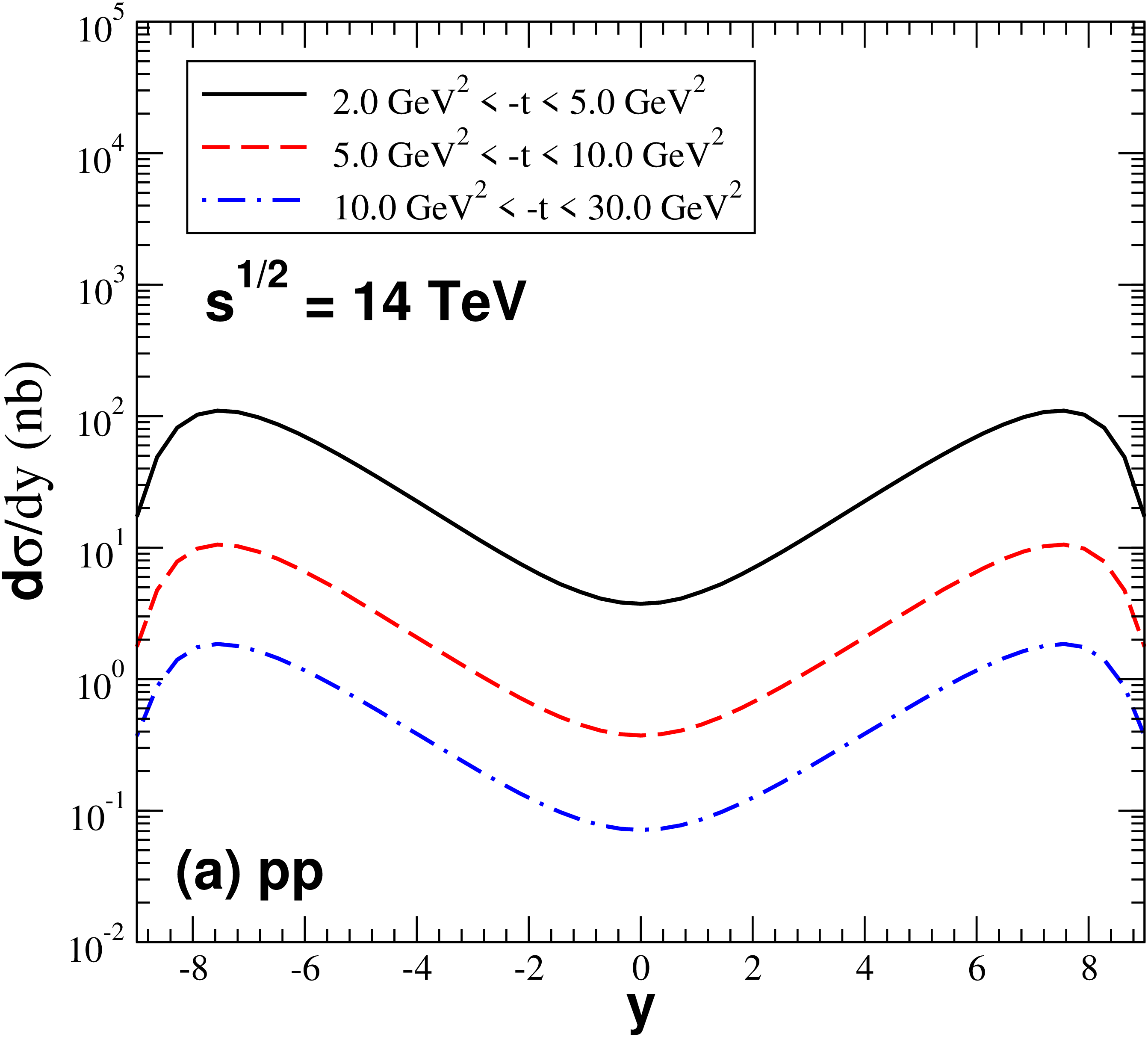} \vspace{0.5cm} &
 \includegraphics[scale=0.15]{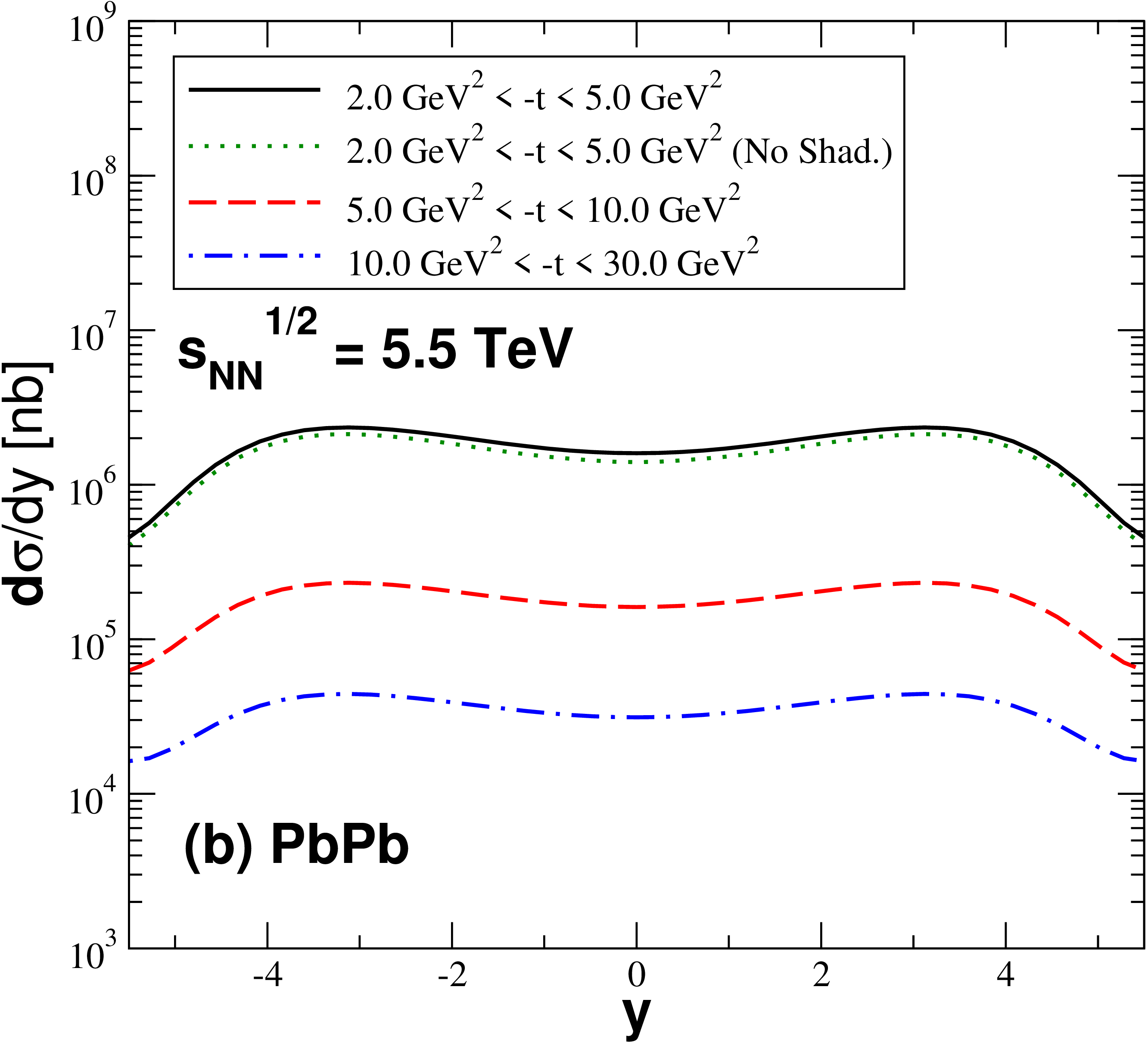}
\end{tabular}
\caption{(Color online) Rapidity distribution for the diffractive $\rho$ photoproduction in (a) $pp$ and (b) $PbPb$ collisions at LHC for distinct $t$-ranges. }
\label{fig2}
\end{figure}

Having fixed the free parameters, let us now calculate the rapidity distribution and total cross sections for diffractive $\rho$ photoproduction in coherent hadron-hadron collisions. The distribution {in} rapidity $y$ of the produced final state can be directly computed from Eq. (\ref{dsigdydt}), by integrating over the squared transverse momentum and using the corresponding  photon spectrum. As the  total cross section for diffractive $J/\Psi$ photoproduction was measured in HERA for different $t$-ranges, we consider three different choices for the limits of integration in the squared momentum transfer. Basically, we assume the same values used in Ref. \cite{vic_sauter}, which define the ranges studied in HERA. This choice is directly associated to the fact that the associated $\gamma p$ data are quite {well described} by our  formalism (See Fig. \ref{fig1}). Moreover, we  integrate over $x_j$  in the range $ 10^{-2} < x_j < 1$, which {means} that we {assume} that the upper limit on the mass of the target dissociation products at LHC is similar to that considered at HERA. 
In Figs. \ref{fig2} (a) and (b) we present our results for the rapidity distribution considering $pp$ collisions at $\sqrt{s} = 14.0$ TeV and $PbPb$ collisions at  $\sqrt{s_{NN}} = 5.5$ TeV, respectively. 
As expected from Fig. \ref{fig1}, the rapidity distribution decreases when we select a range with larger values of $t$. The main differences between our predictions for $pp$ and $PbPb$ collisions are the order of magnitude of the cross sections, which is larger in nuclear collisions due to the higher photon flux (see below), and the prediction of  a plateau in the range $|y| < 2$ in the nuclear case, which is directly associated to the distinct behaviour of the nuclear photon spectrum compared with that of the proton at large photon energies \cite{upcs}.
Moreover, in our calculations for the nuclear case we are assuming  that the nuclear parton distributions are given by the EKS parametrization \cite{eks}.  As the nuclear parton distributions are probed at $x_j \ge 10^{-2}$, our results are sensitive to the magnitude of the antishadowing effect present in the EKS parametrization, which predicts that $f_A(x_j,|t|) > A \times f_p(x_j,|t|)$ and, consequently,  implies an enhancement of the cross section in comparison with the predictions obtained disregarding the nuclear effects. Moreover, the EKS parametrization predicts that these effects diminishes at larger $|t|$. In Fig. \ref{fig2} (b) we present the prediction obtained  disregarding the nuclear effects at the  $2.0 < |t| < 5.0$ range, which we denote No Shad in the plot. In comparison to the EKS prediction at the same $|t|$ range, the No Shad one is almost 10 \% smaller. This difference diminishes at the other $|t|$ ranges and the resulting predictions are almost identical to those obtained using the EKS parametrization. It is important to emphasize that we have verified that similar conclusions are obtained using the most modern parametrization for the nuclear effects proposed in Ref. \cite{eps09}.

In Table \ref{tab1} we present our predictions for the  total cross section considering $pp$ and $PbPb$ collisions. In order to compare with the predictions presented in Ref.  \cite{vic_sauter},  the branching ratios of the vector mesons are not included. As the photon flux is proportional to $Z^2$, {because} the electromagnetic field surrounding the ion is  larger than the proton one due to the coherent action of all protons in the nucleus,  the nuclear cross sections are amplified by a similar factor, which implies very large cross sections for the diffractive $\rho$ photoproduction at large-$t$ in $PbPb$ collisions at LHC.
Considering the {design} luminosities at LHC for $pp$ collisions (${\cal L}_{\mathrm{pp}} = 10^{34}$ cm$^{-2}$s$^{-1}$) and $PbPb$ collisions (${\cal L}_{\mathrm{PbPb}} = 4.2 \times 10^{26}$ cm$^{-2}$s$^{-1}$) we can calculate the production rates (See Table \ref{tab1}). Although the cross section for the diffractive $\rho$ photoproduction at large-$t$ in $AA$ collisions is much larger than in $pp$ collisions, the event rates are higher in the $pp$ mode  due to its larger luminosity. In particular, we predict approximately 7510 events per second in the range 2.0 GeV$^2$ $< |t| <$ 5.0 GeV$^2$ for $pp$ collisions at $\sqrt{s} = 14$ TeV. In contrast, 9 events per second are predicted for $PbPb$ collisions at $\sqrt{s_{NN}} = 5.5$ TeV in the same $t$-range. However, as already emphasized in Ref. \cite{vic_sauter}, for a luminosity above ${\cal L} \ge 10^{33}$ cm$^{-2}$s$^{-1}$, multiple hadron - hadron collisions per bunch crossing are { very likely}, which leads to a relatively large occupancy of the detector channels.
{This} drastically reduces the possibility of {measurement} of coherent
processes at these luminosities. In contrast, at lower luminosities the event pile-up is negligible. Consequently, an estimative considering ${\cal L}_{\mathrm{pp}} = 10^{32}$ cm$^{-2}$s$^{-1}$ should be more realistic. It reduces our predictions for the event rates in $pp$ collisions by a factor $10^2$.

\begin{figure}[t]
\begin{tabular}{cc}
\includegraphics[scale=0.15] {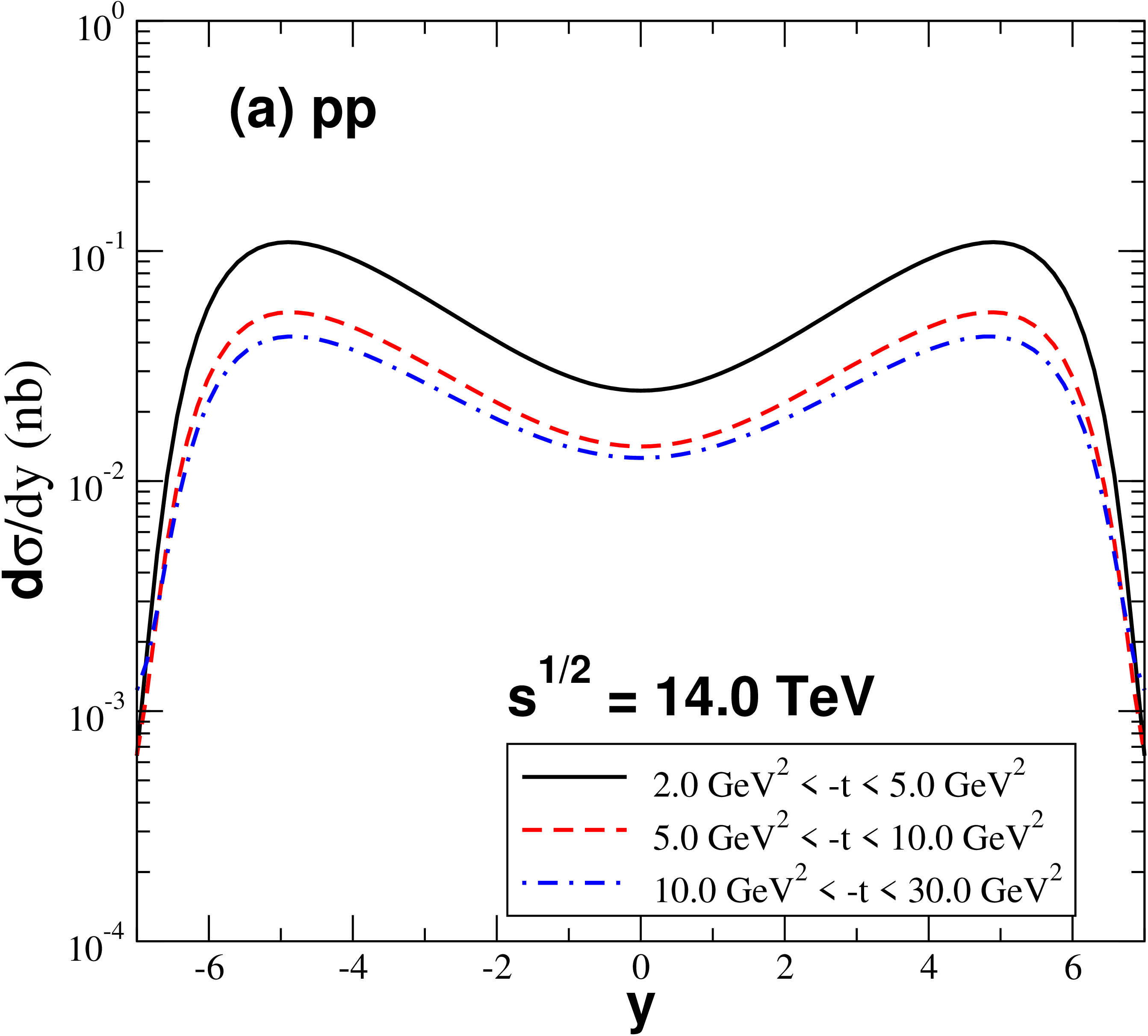} \vspace{0.5cm} &
 \includegraphics[scale=0.15]{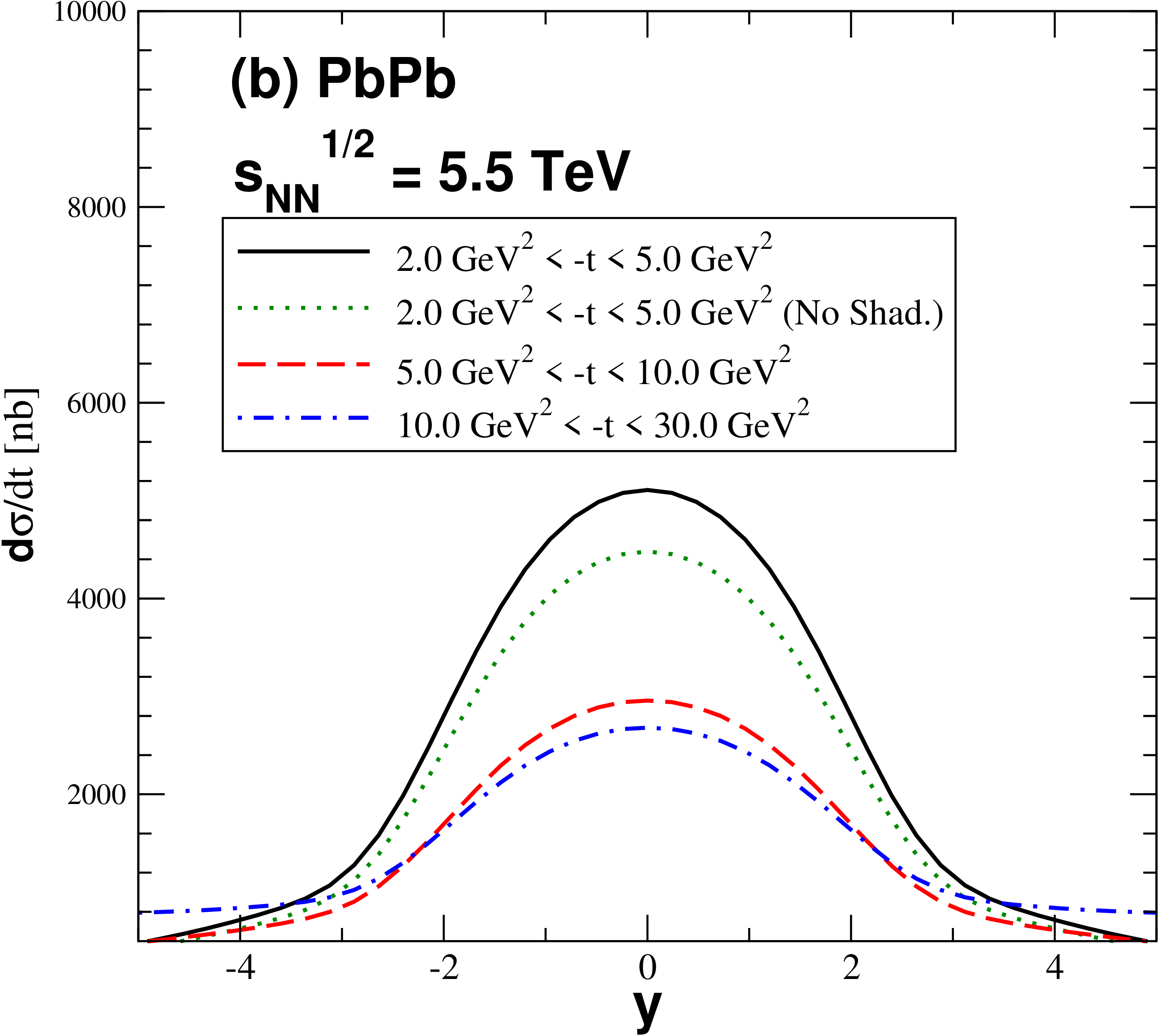}
\end{tabular}
\caption{(Color online) Rapidity distribution for the diffractive $\Upsilon$ photoproduction in (a) $pp$ and (b) $PbPb$ collisions at LHC for distinct $t$-ranges. }
\label{fig3}
\end{figure}

Our analysis can be  extended for $\Upsilon$ production. The only shortcoming is that in this case we cannot compare our predictions for  $d\sigma/dt$ with $ep$ data and, consequently, the choice of the parameters $\gamma$ and $\beta$ in the definition of $\Lambda$ is arbitrary. Following our previous study of the diffractive photoproduction of a heavy vector meson \cite{vic_sauter}, we will assume that $\beta = 1.0$ and $\gamma = 0.25$. Our predictions for the rapidity distribution are presented in Fig. \ref{fig3}. The main difference in comparison to the $\rho$ case is that the distribution is not strongly dependent on the $t$-range considered. It is directly associated to the large mass of the $\Upsilon$, which determines the hard scale  in the process.
Our predictions for the  total cross section considering $pp$ and $PbPb$ collisions are presented in Table \ref{tab1}. In comparison to $\rho$ and $J/\Psi$ production, the total cross sections are at least a factor $10^3$ smaller \cite{note}.  Similarly to the $\rho$ case, the nuclear effects imply an enhancement of the $\Upsilon$ cross section in comparison to the No Shad prediction.

In comparison with the coherent and $t = 0$ case  discussed in \cite{vicmag_rho}, where the target remains intact, the cross section for the $\rho$-production at large-$t$  is a  factor $\ge 10^2$ smaller. In contrast, for $\Upsilon$ production, our predictions for the large-$t$ cross section are similar those presented in \cite{vicmag_ups}. This distinct behaviour is directly associated to the $t$-dependence predicted by the BFKL equation, which implies that the differential cross section $d\sigma / dt$ [See Eq. (\ref{dsdtgq})]  have a steeper behaviour on $t$ for $\rho$ than for $\Upsilon$ production (For a detailed discussion see, e.g. Ref. \cite{vicwer_dm}).



As discussed in detail in \cite{vic_sauter}, a current challenge is the experimental separation of coherent processes. 
The coherent condition { leads to the result
that} the final state has transverse momentum of order of $t$. Consequently, the decay 
{electrons or muons from vector mesons}  have energies of {the order} of $\sqrt{m_{V}^2/4+|t|}$. In the $\rho$ case and low-$t$, such energies are in the lower limits of  acceptances and trigger selection thresholds of the CMS and ATLAS detectors. However, at large-$t$ , the energy of the decay products of $\rho$  is higher than those needed to reach the electromagnetic calorimeter or muon chambers at CMS and ATLAS. Consequently, the detection of the $\rho$ produced in coherent processes is, in principle,  possible  in the CMS and ATLAS detectors. Moreover, 
simulations indicate that the $\Upsilon$ can be identified with a good signal to background ratios when the entire event is reconstructed and a cut is applied on the summed transverse momentum of the event \cite{david} (See also \cite{cox,cms,atlas,f420}). 
Finally, it is important to emphasize that in ALICE, which is designed to handle multiplicities of several thousand particles in a single event, the reconstruction of the low {multiplicity} events associated to coherent processes should not be a problem \cite{alice}.  In particular, the muon arm, which covers the pseudorapidity range $-2.5 > \eta > -4.0$,   should be  capable of reconstructing the vector mesons produced in $pp$ collisions through their dilepton decay channel.



\begin{table}[t]
\begin{center}
\begin{tabular}{||c|c|r@{.}l|r@{.}l||}
\hline
\hline
Meson & $t$ range  & \multicolumn{2}{c|}{$pp$ } & \multicolumn{2}{c||}{$PbPb$ }  \\
\hline
\hline
$\rho$ & $2.0 < |t| < 5.0$  &  751&0 nb (7510.0) & 20&0 mb (8.4)  \\
\hline
& $5.0 < |t| < 10.0$  &  71&0 nb (710.0) & 2&2 mb  (0.9) \\
\hline
& $10.0 < |t| < 30.0$  &  12&0 nb (120.0)& 0&4 mb  (0.17)\\
\hline
\hline
$\Upsilon$ & $2.0 < |t| < 5.0$  &  0&8 nb (8.0) & 0&26 mb (0.1)  \\
\hline
& $5.0 < |t| < 10.0$  &  0&4 nb (4.0) & 0&17 mb  (0.07) \\
\hline
& $10.0 < |t| < 30.0$  &  0&3 nb (3.0)& 0&16 mb  (0.06)\\
\hline
\hline
\end{tabular}
\end{center}
\caption{The integrated cross section (event rates/second)  for the diffractive vector meson photoproduction at large momentum transfer in $pp$ and $PbPb$ collisions at LHC. The squared momentum transfer, $t$, is given in GeV$^2$.} 
\label{tab1}
\end{table}

In summary, in this paper we have extended the study of the diffractive photoproduction of vector mesons with hadron dissociation in coherent interactions for the production of $\rho$ and $\Upsilon$.  The corresponding cross sections and event/rates are large, which implies that the experimental analysis of these processes should be feasible at LHC. 

\section*{Acknowledgements}
 This work was partially financed by the Brazilian funding agencies CNPq, CAPES and FAPERGS.




\begin{thebibliography}{99}
\bibitem{note}
{Due to a misprint, the integrated cross section for the diffractive $J/\Psi$ photoproduction in $PbPb$ collisions presented in Table I from \cite{vic_sauter}   should be divided by a factor 10.}
\end{thebibliography}
\end{document}